\documentclass[12pt,a4paper,dvips]{revtex4}
\usepackage[ansinew,latin1]{inputenc}
\usepackage[T1]{fontenc}
\usepackage{graphicx}
\usepackage{amsmath}
\usepackage{amssymb}

\usepackage{latexsym}
\usepackage{feynmf}     

\def\beq{\begin{equation}}
\def\eeq{\end{equation}}
\def\bitem{\begin{itemize}}
\def\eitem{\end{itemize}}
\def\bear{\begin{array}}
\def\ear{\end{array}}

\def\munu{{\mu\nu}}

\def\d{\partial}

\def\m{\hat{m}}

\begin{document}

\begin{titlepage}
\begin{center} {\LARGE \bf Particles and fields within a unification scheme
\vspace{0.2in} \\}\vspace*{1cm}
{\bf Francesco Cianfrani$^*$, Valentino Lacquaniti$^{\dag*\S}$, Giovanni Montani$^{*\ddag}$}\\
\vspace*{1cm}
$^\dag$ Physics Department  "`E.Amaldi "`,  University of Rome, ``Roma Tre'',\\
Via della Vasca Navale 84, I-00146, Rome , Italy.

\vspace*{0.5cm}

$^*$ ICRA-International Center for Relativistic Astrophysics,,\\ Dipartimento di Fisica (G9),
University of Rome, ``Sapienza'',\\ Piazzale Aldo Moro 5, 00185 Rome, Italy.

\vspace*{0.5cm}

$^\ddag$ ENEA C.R. Frascati (U.T.S. Fusione), via Enrico Fermi 45, 00044 Frascati, Rome, Italy.\\ 
ICRANet C. C. Pescara, Piazzale della Repubblica, 10, 65100 Pescara, Italy.

\vspace*{0.5cm}

$^\S$ LAPTH -9, Chemin de Bellevue BP 110 74941 Annecy Le Vieux Cedex, France.

\vspace*{0.5cm}

e-mail: lacquaniti@fis.uniroma3.it, francesco.cianfrani@icra.it, montani@icra.it\\
\vspace*{1.8cm}

 \vspace*{1cm} \\

\vspace*{1cm}

{\bf  Abstract\\} \end{center}

We discuss properties of particles and fields in a multi-dimensional space-time, where the geometrization of gauge interactions can be performed. For instance, in a 5-dimensional Kaluza-Klein manifold we argue that the motion of charged spinning bodies is obtain in a Papapetrou-like formulation. As far as spinors are concerned, we outline how the gauge coupling can be recognized by a proper dependence on extra-coordinates and by the dimensional reduction procedure. 

\end{titlepage}

\section{Kaluza-Klein 5D Model}
In 5D Kaluza-Klein (KK) model \cite{kaluzaklein}, given a 5D metrics $J_{AB}$, the  generalization of GR is  provided, with obvious formalism, as follows: 
\beq
ds^2_{5}=J_{AB}dx^A dx^B \quad
S_5=-\frac{1}{16\pi G_{(5)}}\int\!\!d^5\!x \sqrt{J}\,\, {}^5\!R \quad A,B=0,1,2,3,5
\label{s5}
\eeq
There are three assumptions: i) we consider  a manifold  $\mathcal{M}^5=\mathcal{M}^4\otimes S^1$, where $\mathcal{M}^4$ is the ordinary space-time and $S^1$ is a space-like loop; ii) all components of the metric tensor are not depending on the extra coordinate $x^5$ (cylindricity hypothesis); iii) the $J_{55}$ component of the metrics is  a constant ( $J_{55}=-1$).
The model is not invariant with respect to general  diffeomorphism but it is now  invariant with respect to  the transformations 
\beq
x^5=x^{5'}+ek\Psi(x^{\mu'}) \quad\quad
x^{\mu}=x^{\mu}(x^{\nu'}),
\label{kktrasf}
\eeq 
where $\Psi$ is a scalar function and $ek$ an appropriate dimensional factor. 
In agreement with the above rules,  $J_{AB}$ admits the following representation:
\beq
\left\{
\begin{array}{l}
J_{55}=-1 \\
J_{5\mu}=-ekA_{\mu}   \\
J_{\munu}=g_{\munu}-(ek)^2 A_{\mu}A_{\nu}
\end{array}
\right.
\rightarrow
J_{AB}{}\Rightarrow{}\left(
\begin{array}{ccc}
g_{\munu}-(ek)^2A_{\mu}A_{\nu} & -ekA_{\mu} \\
                 &            \\
-ekA_{\mu}     & -1
\end{array}
\right)
\label{splittingmetrics}
\eeq 
 With respect to KK diffeomorphism (\ref{kktrasf}), $g_{\munu}$ is a 4D tensor and $A_{\mu}$ is a 4D gauge vector :
\beq
g_{\munu}'=g_{\rho\sigma}\frac{\d x^{\rho}}{\d x^{{\mu}'}}\frac{\d x^{\sigma}}{\d x^{{\nu}'}}  \quad\quad
A_{\mu}'=A_{\nu}\frac{\d x^{\nu}}{\d x^{{\mu}'}}+\frac{\d\Psi}{\d x^{{\mu}'}}
\eeq
In agreement with equations (\ref{splittingmetrics}),  by using the 5-bein  projection (\cite{cmm05})  for  the splitting of ${}^5R$ 
and the identity  $\sqrt{J}=\phi\sqrt{-g}$, after KK splitting the Action reads:
\beq
S_5=-\frac{1}{16\pi {G_5}'} \int\!d^4\!x \sqrt{-g}\, (\phi R+\frac{1}{4}(ek)^2F_{\munu}F^{\munu} )
\label{kkaction}
\eeq
where $R$ is the usual curvature scalar , $F_{\munu}$ the Faraday tensor and ${{G_5}'}^{-1}={G_5}^{-1}\int\!dx^5$.
If we set $G={G_5}'$ and $\frac{4G}{c^4}=(ek)^2$
we recognize that KK theory reproduces, in vacuum,  the Einstein-Maxwell theory. The notably feature is the identification of gauge symmetry with a particular case of diffeomorphism invariance. However, an expensive prize has to be payed; the 5D  General Covariance is broken ( while the 4D still holds, so that GR is safe ).  The same happens for the 5D Principle of Equivalence: due to the law of transformations  of metrics components  ( that does not admit non-linear transformations as far as $x^5$ is concerned ),  there is no way to find a local transformations which could bring us to a Minkowsky 5D space ( except for the case $A_{\mu}=0$,  which is not interesting to us ).
\paragraph{The problem of motion}
 By assuming that the motion of a free-falling 5D test particle is driven by the  geodesic Action as in GR, we consider
\beq
S^{geod}_5=\int ds_5  \quad\quad ds_5^2=ds^2- (ekA_{\mu} dx^{\mu}+d x^5)^2
\label{act5}
\eeq
where the line element  is split in agreement with (\ref{splittingmetrics}). 
The variational procedure yields a set of five equation; by writing them as functions of $ds$ we get:
\beq
\frac{d}{ds}w_5=0 \quad\quad
\frac{D}{Ds}u^{\mu}=ekF^{\munu}u_{\nu}(\frac{w_5}{\alpha})
\label{sistema1}
\eeq
The first equation sets the existence of a constant of motion , $w_5$ ( namely, the fifth component of the 5-velocity $w_A=J_{AB}\frac{dx^A}{ds_5}$ ), that is also a scalar as it could be shown by a direct calculus.  Other four describe the motion of a particle with velocity $u^{\mu}=\frac{dx^{\mu}}{ds}$, where $\frac{D}{Ds}$ is  the usual derivative along the path and $\alpha$ is  : $\alpha=\frac{d s}{ds_5}=\sqrt{1+w^2_5}$.
We recover the Lorentzian electrodynamics once we define  the charge-mass ratio :
\beq
\frac{q}{mc^2}=ek\frac{w_5}{\sqrt{1+w^2_5}} 
\label{charge22}
\eeq
However, the right member of the above equation is upper bounded, and this yields, recalling the value of $ek$, to a bound on the charge-mass ratio:
\beq
\frac{q^2}{4Gm^2}< 1
\eeq
Elementary particles does not satisfy this  bounding \footnote{If we consider the electron, for instance, we have $\frac{e^2}{4Gm_e^2}\sim (10^{43})$}, therefore the geodesic approach  gives an unphysical description for the test particle motion. This is a valuable example of what a puzzle is the problem of matter coupling in KK theories \cite{matterproblem}. The $q/m$  problem is strictly linked to the problem of the huge massive particles generated by KK modes. 
Indeed, considering the geodesic identity for a 5D test particle, namely $J^{AB}w_Aw_B=1$,
by introducing a mass parameter $\m$ and defining linear momentum $p_A=\m w_A$, we gain the 5D dispersion relation
$
J^{AB}p_Ap_B=\m^2.
\label{5ddisp}
$
Now, as a toy model, we  quantize this relation and we get a 5D Klein-Gordon equation. The associated Lagrangian density for a complex scalar field $\zeta$ reads: $\mathcal{L}=J^{AB}\d_A\zeta(\d_B\zeta)^+-\m^2\zeta\zeta^+$.
We assume a dependence on $x^5$ only through a phase factor: $\zeta(x^{\mu},x^5)=\eta(x^{\mu})e^{ip_5x^5}$,
where $p_5$ is scalar and conserved, thus our field transforms as a U(1) gauge field: $\zeta=\zeta'e^{i(ekp_5)\Psi}$.
The reduction of the Lagrangian yields :
\beq
\mathcal{L}=g^{\munu}(-i\d_{\mu}-p_5ekA_{\mu})\eta[(-i\d_{\nu}-p_5ekA_{\nu})\eta]^+-(\m^2+{p_5^2})\eta\eta^+
\label{51}
\eeq
The reduced field $\eta$ acquires  a charge $ekp_5$ and a mass term $m^2=(\m^2+\frac{p_5^2}{\phi^2})$; the ratio $q/\m$ we get fits with the result previously obtained for a test particle, but is clear now that  $\m$ does not represent the correct rest mass for the particle. By requiring the compactness of the fifth dimension we get  the quantization of $p_5$, and so on of the charge. Thus, the discretization of $p_5$ gives rise to a tower of modes for the mass term $m$;  fixing the minimum value of $p_5$ via the elementary charge $e$ we get the extra dimension size below our observational bound, but, at the same way, we get huge massive modes beyond the Planck scale. Therefore, the puzzle of the charge-mass ratio is strictly linked to the problem of massive modes. 
\paragraph{Papapetrou approach}
Granted the unphysical outcome of the geodesic procedure, we purpose a new approach: our aim is to deal directly with the generic 5D matter tensor $T^{AB}$  and  thereafter address the problem of the localization of the test particle via the multipole expansion \emph{a la Papapetrou} \cite{pap}.
We assume that is possible to state a conservation laws and a condition of consistency with cylindricity hypothesis:
$$
D_A T^{AB}=0 \quad\quad
\d_5T^{AB}=0
$$
The dimensional reduction of the above set yields:
$$
eq.5) \rightarrow \nabla_{\mu} j^{\mu}=0 \quad
eq. \mu) \rightarrow \nabla_{\rho}T^{\mu\rho}=F^{\mu\rho} j_{\rho}
$$
where we  put $j_{\mu}=-ekT_5^{\mu}$. Then we have a $4D$ conserved current, thus this model   reproduces the Lorentzian electrodynamics ;  the current is only defined  in terms of components of the 5D matter tensor, without any connection with the kinematics of the matter ( no link between charge and the fifth component of the velocity). At this stage any definition of mass has not  been employed and therefore we expect  not to have  any kind of restriction on the value of the current.  To deal with the localization of the particle let us consider now the multipole expansion.
The Papapetrou procedure consists in many steps, so we just sketch it. Given the above set of equations we first write explicitly the dependence on Christoffel;  we also take the derivative of $x^{\nu}\sqrt{g}T^{\mu\lambda}$, $x^{\nu}\sqrt{g}T_5^{\mu}$ and we get an auxiliary set of equations.
We integrate our equations over the 3-space  and perform the multipole expansion via the hypothesis of localization of the matter tensor: given a world line $X^{\mu}$ in the 4D space, we assume that the  matter tensor is peaked in a thin tube centered around the trajectory $X^{\mu}$ and negligible  outside. Then we perform a  Taylor expansion of center $X^{\mu}$ and  consider only the lowest order; indeed we assume that at the lowest order components of tensor are proportional to a 4D Dirac delta. Equations of motion read:
\beq
\frac{D}{Ds}(mu^{\mu})=qF^{\mu\rho}u_{\rho} \quad\quad \frac{dq}{ds}=0
\label{moto1}
\eeq
where $ds^2=g_{\munu}dX^{\mu}dX^{\nu}$, $u^{\mu}=\frac{dX^{\mu}}{ds}$. Christoffel are now included in $\frac{D}{Ds}$ and
\begin{eqnarray*}
&{}& m=\frac{1}{u^0}\int\!\!\!d^3x\,\sqrt{g}T^{00} \quad\Rightarrow \quad \sqrt{g}T^{\mu\nu}=\int\!\!\!ds\,m\,\delta^4(x-X)u^{\mu}u^{\nu} \\
&{}& q=\int\!\!\!d^3x\,\sqrt{g}T_5^0\quad\quad\quad \Rightarrow\quad \sqrt{g}T_5^{\mu}=\int\!\!\!ds\,q\,\delta^4(x-X)u^{\mu} 
\end{eqnarray*}
Quantities $m$, $q$,  result to be scalar objects; $q$ is a constant of motion while the conservation of $m$ arises from the subsidiary condition  $u_\mu\frac{Du^{\mu}}{Ds}=0$. 
Thus we recover the Lorentz equation for a test particle. The difference with respect to the equation provided by the geodesic approach relies in the definition of couplings $m$ and $q$; while previously they were linked each other, via the quantity $w_5$, and therefore their ratio were upper bounded,
now we are able to define charge and mass in terms of independent degrees of freedom, namely $T^{00}$ and $T_5^0$, and hence the $q/m$ is not affected by any bound. At the same time we can show that there is a suppression of the huge massive modes.
Indeed, the set of equations (\ref{moto1}) admits a Lagrangian description: given a set of Lagrangian coordinates $y^{A}=(x^{\mu},l)$ the motion is described by the Action
\beq
S=-\int\!\!\!m\,d\,s+q(A_{\mu}dx^{\mu}+dl)
\label{azmoto}
\eeq
After calculating Lagrangian, Hamiltonian and momenta we have:
$$
P_AP^A=m^2-q^2 \quad \Rightarrow \quad  P_{\mu}P^{\mu}=m^2
$$
where
$ P_5=q $, $ \Pi_i=mu_i $, $ P_i=\Pi_i+qA_i $.
Therefore now in the resulting 5D Klein-Gordon equation we have the  quantity $-q^2$  that, when we repeat the quantization procedure for a complex scalar field, acts as a counter-term ruling out the huge massive modes. Hence we  can keep working with a compactified dimension without loss of physical meaning. This procedure shows  that in the 5D ambient the motion of the test particle is not represented by the 5D geodesic world line; the Action (\ref{act5}) leads to the geodesic trajectory but it does not fit with the true Action for the motion (\ref{azmoto}) provided by the lowest order of multipole expansion.
While in 4D they coincide this is no more true in the 5D KK model, and the reason relies in the violation of the PE, due to the cylindricity condition.  It is worth nothing that a key point of this procedure is the extension of the cylindricity to the matter tensor; this fact is responsible of the localization of the tensor as a function of a 4D Dirac delta, rather than a 5D delta. Indeed, the procedure of localization of the particle involves just the usual 4D dimensions; as a consequence of this now $P_5$ is just the conjugate momentum to the fifth coordinate but it has no link with the fifth lagrangian velocity, while the 4D conjugate momenta $P_{\mu}$  is proportional to the velocity through the correct rest mass, as we correctly expect from the right physical viewpoint.

\section{Spinning particles in Kaluza-Klein theory}
 
Let us now consider the next order of a multipole expansion, {\it i.e.} the pole-dipole case. A description of the dynamics is given by Papapetrou equations \cite{1}, whose form is as follows

\begin{equation*}
\left\{\begin{array}{c}\frac{D}{{}^{(5)}\!Ds}{}^{(5)}\!P^{A}=\frac{1}{2}{}^{(5)}\!R_{BCD}^{\phantom1\phantom2\phantom3 A} \Sigma^{BC}{}^{(5)}\!u^{D}\quad\\
\frac{D}{{}^{(5)}\!Ds}\Sigma^{AB}={}^{(5)}\!P^{A}{}^{(5)}\!u^{B}-{}^{(5)}\!P^{B}{}^{(5)}\!u^{A}\\
{}^{(5)}\!P^{A}={}^{(5)}\!m{}^{(5)}\!u^{A}-\frac{D\Sigma^{AB}}{{}^{(5)}\!Ds}{}^{(5)}\!u_{B}\quad\\
\Sigma^{AB}{}^{(5)}\!u_{A}=0\label{pe5}\qquad\qquad\qquad\qquad\end{array}\right.,
\end{equation*}

$\Sigma^{AB}$ and ${}^{(5)}\!P^{A}$ being the 5-spin tensor and the generalize 5-momentum, respectively, while the last relation is the Pirani consistency condition \cite{P56}. 
In view of performing the dimensional reduction down to $V^4$, we identify into $\Sigma^{AB}$ some quantities, $S^{\mu\nu}=\Sigma^{\mu\nu}$ and $S_\mu=\Sigma_{5\mu}$, whose behavior is that of 4-dimensional quantities. 
Within this scheme, the splitting of the Pirani condition $\Sigma^{AB}{}^{(5)}\!u_{A}=0$ gives for $A=\mu$
\begin{equation*}
\alpha\{S^{\nu\mu}u_{\nu}+S^{\mu}u_{5}\}=0,
\end{equation*}
while for $A=5$ no independent condition comes out.\\
As far as the definition of ${}^{(5)}\!P^{A}$ is concerned, it provides the following relations, after evaluating 5-dimensional derivatives of $S^{\mu\nu}$ and $S_\mu$,
\begin{eqnarray}
{}^{(5)}\!P^{\mu}=\alpha^2[P^{\mu}+u_{5}\frac{DS^{\mu}}{Ds}-ekF_{\rho\nu}u^{\rho}S^{\nu\mu}u_{5}]=\alpha^2\widetilde{P}^{\mu}\\
{}^{(5)}\!P_{5}=\alpha^2[mu_{5}-u_{\nu}\frac{DS^{\nu}}{Ds}+ekF_{\rho\nu}u^{\rho}S^{\nu\mu}u_{\mu}]=\alpha^2\widetilde{P}_{5}.
\end{eqnarray}
Hence, the equations for the precession of the spin tensor can be re-written for $A=\mu$ and $B=\nu$ as
\begin{equation}
\frac{DS^{\mu\nu}}{Ds}=\alpha^2[\widetilde{P}^{\mu}u^{\nu}-\widetilde{P}^{\nu}u^{\mu}]+\frac{1}{2}ekF^{\mu}_{\phantom1\rho}(u^{\rho}S^{\nu}+S^{\rho\nu}u_{5})-\frac{1}{2}ekF^{\nu}_{\phantom1\rho}(u^{\rho}S^{\mu}+S^{\rho\mu}u_{5}),
\end{equation}
while the case $A=\mu$ and $B=5$ does not provide an independent condition. This clearly indicates that the additional degrees of freedom a 5-spin tensor has does not add any dynamical information on the 4-spin precession.\\
Finally, the dynamics of particles center of mass is inferred from the splitting of equations for the momentum.  
This procedure gives, for $A=5$, the following conservation law 
\begin{equation*}  
\frac{D}{Ds}(\alpha^{2}\widetilde{P}_{5}+\frac{1}{4}ekF_{\mu\nu}S^{\mu\nu})=\frac{D}{Ds}q=0.
\end{equation*}
The quantity $q$, preserved during the motion, has a natural interpretation in terms of the electric charge. This interpretation is confirmed by the splitting of the equation for ${}^{(5)}\!P^{\mu}$, which gives, as soon as components of the Riemann tensor are evaluated and a new momentum $\hat{P}^{\mu}=\alpha^{2}\widetilde{P}^{\mu}+\frac{1}{2}ekF^{\mu}_{\phantom1\rho}S^{\rho}$ is introduced,
\begin{eqnarray*}
\frac{D}{Ds}\hat{P}^{\mu}=\frac{1}{2}R_{\alpha\beta\gamma}^{\phantom1\phantom2\phantom3\mu}S^{\alpha\beta}u^{\gamma}+qF^{\mu}_{\phantom1\nu}u^{\nu}+\frac{1}{2}\nabla^{\mu}F^{\nu\rho}M_{\nu\rho},
\end{eqnarray*}
$M^{\mu\nu}$ being 
\begin{equation}
M^{\mu\nu}=\frac{1}{2}ek(S^{\mu\nu}u_{5}+u^{\mu}S^{\nu}-u^{\nu}S^{\mu}).
\end{equation}
By collecting together all previous results, we get the system
\begin{equation}
\left\{\begin{array}{c}\frac{D}{Ds}\hat{P}^{\mu}=\frac{1}{2}R_{\alpha\beta\gamma}^{\phantom1\phantom2\phantom3\mu}S^{\alpha\beta}u^{\gamma}+qF^{\mu}_{\phantom1\nu}u^{\nu}+\frac{1}{2}\nabla^{\mu}F^{\nu\rho}M_{\nu\rho}.\\
\frac{DS^{\mu\nu}}{Ds}=\hat{P}^{\mu}u^{\nu}-\hat{P}^{\nu}u^{\mu}+F^{\mu}_{\phantom1\rho}M^{\rho\nu}-F^{\nu}_{\phantom1\rho}M^{\rho\mu}\\
\hat{P}^{\mu}=\alpha^{2}P^{\mu}+u_{5}\frac{DS^{\mu}}{Ds}-ekF_{\rho\nu}u^{\rho}S^{\nu\mu}u_{5}+\frac{1}{2}ekF^{\mu}_{\phantom1\rho}S^{\rho}\\
S^{\nu\mu}u_{\nu}+S^{\mu}u_{5}=0\end{array}\right.,\label{splpap}
\end{equation}
which coincide precisely with Dixon-Souriau equation \cite{5,4}, as soon as $M^{\mu\nu}$ is identified with the electro-magnetic moment of the body under investigation. This result shows that the geometrization of the electro-magnetic fields in a KK background does not modify the dynamics of the moving object, up to the dipole order.\\
Within the definition of the electromagnetic tensor, the following identifications stand
i) $S^{\mu\nu}$ 4-spin tensor, ii) $S^{\mu}$ electric dipole. Therefore, the extra-components of the 5-spin tensor ($ \Sigma_{5\mu}$) describe a non-vanishing electric dipole moment.\\ 
We point out that, in agreement with results of the previous section, we expect the equations (\ref{splpap}) to be extensible to elementary particles by performing the multipole expansion appropriate to the KK hypothesis.

\section{MATTER FIELDS IN A KALUZA-KLEIN SPACE-TIME}
Matter fields have to be introduced according with symmetries of the KK space-time. In particular, an additional component of the momentum should come out and its physical meaning is going to be clarified \cite{M,cmm05}.\\   
In this respect, let us consider an empty space-time $M^{4}\otimes
S^{1}$, $M^4$ being Minkowski space-time, on which some matter fields $\varphi_{r}$ are put as perturbations. Their dynamics is described by a Lagrangian density
\begin{equation}
\Lambda=\Lambda(\varphi_{r};\partial_{A}\varphi_{r}),
\end{equation}
which reflects the restricted General Covariance, hence it is invariant under the infinitesimal coordinates
transformations (\ref{kktrasf}).\\ 
Under global transformations, the behavior of matter fields is the following one
\begin{equation}
\varphi'_{r}=\varphi_{r}+\delta\varphi_{r},\qquad
\delta\varphi_{r}=\partial_{A}\varphi_{r}u^{A}_{\bar{B}}\delta\omega^{\bar{B}},
\end{equation}
while for the Lagrangian density we have
\begin{eqnarray*}
\delta\Lambda=(\partial_{A}\Lambda)
u^{A}_{\bar{B}}\delta\omega^{\bar{B}}=\frac{
\partial\Lambda}{\partial\varphi_{r}}\delta\varphi_{r}+\frac{
\partial\Lambda}{\partial(\partial_{A}\varphi_{r})}\delta(\partial_{A}\varphi_{r})
=\partial_{A}\bigg(\frac{
\partial\Lambda}{\partial(\partial_{A}\varphi_{r})}\delta\varphi_{r}\bigg)-\bigg[\partial_{A}\bigg(\frac{
\partial\Lambda}{\partial(\partial_{A}\varphi_{r})}\bigg)-\frac{
\partial\Lambda}{\partial\varphi_{r}}\bigg]\delta\varphi_{r}.
\end{eqnarray*}
Hence, by virtue of Euler-Lagrange equations in curved space, the continuity equation below is obtained
\begin{equation}
\label{a2}\nabla_{A}\bigg(\frac{
\partial\Lambda}{\partial(\partial_{A}\varphi_{r})}(\partial_{C}\varphi_{r})u^{C}_{\bar{B}}-\Lambda
u^{A}_{\bar{B}}\bigg)=0,
\end{equation}
whose associated conserved quantities read as
\begin{equation}
\label{b2}P_{\bar{A}}=\int_{E^{3}\otimes S^{1}}
\sqrt{-\gamma}[\Pi_{r}(\partial_{B}\varphi_{r})u^{B}_{\bar{A}}-\Lambda
u^{0}_{\bar{A}}]d^{3}xdy,
\end{equation}
where $E^{3}$ is euclidean three-dimensional space and $\Pi_{r}$
the fields conjugated to $\varphi_{r}$.

The expressions (\ref{b2}) are components of the 5-momentum and, as soon as a phase dependence on the extra-coordinate is taken for $\varphi_{r}$, {\it i.e.}
\begin{equation}
\varphi_{r}=\frac{1}{\sqrt{L}}e^{-iy/L}\phi_{s}(x^{\mu})\qquad
\Pi_{r}=\frac{1}{\sqrt{L}}\pi_{s}(x^{\mu})e^{iy},\label{c2}
\end{equation}
they can be rewritten as
\begin{eqnarray}
Q_{\mu}=P_{\mu}=\int_{E^{3}}[\pi_{r}\partial_{\mu}\phi_{r}-\Lambda\delta^{0}_{\mu}]d^{3}x\qquad
\label{e2}Q=\frac{i}{L}\int_{E^{3}}(\pi_{r}\phi_{s})d^{3}x.
\end{eqnarray}
It is worth nothing that the first terms are components of
4-momentum, whose conservation is a consequence of the invariance under 4-dimensional translations. The last component coincides with the charge, associated with the $U(1)$ symmetry.\\
Therefore, as soon as matter fields are introduced, gauge transformations are going to be identified with 
extra-dimensional translations and, in fact, by the proper phase dependence, a U(1) transformation law is inferred 
\begin{equation}
\label{g2}\phi'_{r}=\phi_{r}+i\delta\omega\phi_{r}.
\end{equation}

\subsection{Fermions in a Kaluza-Klein framework}
The introduction of spinors in a KK scenario is suitable to results of the previous section, since a phase dependence for them implies that an y-dependence is not observable, thus it realizes an extension of the cilindricity condition. In this respect, we assume for 5-spinor $\Psi$
\begin{equation}
\Psi(x,y)=e^{iy/L}\psi(x).
\end{equation}
The presence of an additional dimension implies to find out an additional Dirac matrix. The choice is fixed by the underlying algebra being a Clifford one, in particular by standard anti-commutation relations, hence in 5-bein components the fifth matrix coincides with the chirality operator $\gamma_5$.\\
Since we expect, in a scenario with a compactified dimension, the multidimensional Lorentz invariance to be broken, spinor connections different from Riemannian ones are going to be taken. In particular, if they are fixed as
\begin{eqnarray}\left\{\begin{array}{cc}
\Gamma_{(\mu)}=^4\Gamma_{(\mu)}\equiv- \frac{1}{4}\gamma^{(\rho)}
\gamma^{(\sigma)}R_{(\sigma)(\rho)(\mu)}\\{} \,\,\Gamma_{(5)}=\frac{i}{L}\textbf{I}, \label{114}
\end{array}\right.\end{eqnarray}the 4-dimensional Lorentz invariance is preserved. Moreover, the dimensional reduction of the Dirac Lagrangian density
\begin{equation}
 \Lambda=-\frac{i\hbar c}{2}\bar{\chi}
   \gamma^{(A)}D_{(A)}\chi+\frac{i\hbar c}{2}
   (D_{(A)}\bar{\chi})\gamma^{(A)}\chi+
   imc^2\bar{\chi}\chi
\end{equation}
gives the 4-dimensional one, in presence of the right $U(1)$ gauge interaction and without the appearance of KK mass terms, {\it i.e.}
\begin{eqnarray}
    S=\frac{1}{c}\int\sqrt{-g}\Phi\big[
    -\frac{i\hbar c}{2}\bar{\psi}\gamma^\mu
    D_\mu\psi+\frac{i\hbar c}{2}(D_\mu\bar{\psi})
    \gamma^\mu\psi-\frac{2\pi ek\hbar c}{L}
    A_\mu\bar{\psi}\gamma^\mu\psi+
    imc^2\bar{\psi}\psi\big]\,d^4x\nonumber\\ \label{121}
\end{eqnarray}
From the full Lagrangian density, it can be recognized that the ordinary action, with the terms
$\frac{1}{16\pi}F_{\mu\nu}F^{\mu\nu}$ and
$eA_\mu\bar{\psi}\gamma^\mu\psi$, arises by imposing the following conditions 
\begin{equation}
\frac{e^2k^2c^4}{4G}=1
\qquad
\frac{2\pi ek\hbar c}{L}=e,       \label{123}
\end{equation}
which provide $k=\frac{\sqrt{4G}}{ec^2}$ and an estimate $L=2\pi\sqrt{4G}\frac{\hbar}{ec}=
4.75\,10^{-31}cm$ for the length of the extra-coordinate.

\end{document}